\documentclass[aps,twocolumn,groupedaddress,showpacs]{revtex4-1}

\usepackage{graphicx,bm,bbm,amsmath,lmodern}

\newcommand{\aver}[2]{\ensuremath{\left\langle#1\right\rangle_{#2}}}

\newcommand{\pd}[2]{\ensuremath{\frac{\partial #1}{\partial #2}}}
\newcommand{\pr}[1]{\mathrm{Pr}\left\{#1\right\}}
\newcommand{\tr}{{\ensuremath{\rm tr}\,}}
\newcommand{\capa}{\ensuremath{\mathcal{C}}}
\newcommand{\var}{{\ensuremath{\rm Var}}}
\newcommand{\corr}{{\ensuremath{\rm corr}}}
\newcommand{\T}{{\ensuremath{\sf T}}}
\newcommand{\erf}{{\ensuremath{\rm erf}}}
\newcommand{\sgn}{{\ensuremath{\rm sgn}}}

\begin{document}

\title{Information transfer with small-amplitude signals}

\author{Lubomir Kostal}
\email[]{kostal@biomed.cas.cz}
\author{Petr Lansky}

\affiliation{Institute of Physiology AS CR, v.v.i., Videnska 1083, 142
20 Prague 4, Czech Republic}

\date{\today\ (Fig.1 improved); 
May 18, 2010: Published  in Phys. Rev. E, 
DOI:{\tt 10.1103/PhysRevE.81.050901}}

\begin{abstract}
We study the optimality conditions of information transfer in systems
with memory in the low signal-to-noise ratio regime of vanishing input
amplitude.  We find that the optimal mutual 
information is represented by a
maximum-variance of the signal time course, with correlation structure
determined by the Fisher information matrix.  We  provide
illustration of the method on a simple biologically-inspired model of
electro-sensory neuron.  Our general results apply also to the study of
information transfer in single neurons subject to weak stimulation,
with implications to the problem of coding efficiency in biological
systems. 
\end{abstract}

\maketitle

Theoretical approach to the problem of information  processing in
biological (neuronal) systems has received significant attention over
the past few decades \cite{r:dayan, r:rieke}, with information theory
\cite{r:ash, r:cover, r:verdurev} providing the fundamental framework
\cite{r:brunelnadal, r:borstrev, r:bulsarapre, r:nemenman, r:stein-bj,
r:sompolinsky-cor,  r:victor}.  Of particular interest are the
optimality conditions under which the information between stimuli and
responses is maximized \cite{r:atick, r:bialek90, r:laughlin,
r:mcdonnell-cap, r:koslanrosp08plos}, leading to the idea of
\emph{efficient coding hypothesis} \cite{r:barlow}.
Due to non-linear nature of
information-theoretic measures, explicitly formulated  optimality
conditions are relatively rare \cite{r:cover, r:davis,
r:verdu-capcost}, nevertheless numerical methods exploiting properties
of mutual information are available \cite{r:cover, r:ikeda,
r:mcdonnell-cap}.  Recently, the asymptotic relation between mutual
information and Fisher information \cite{r:bernardo79, r:rissanen} has
been employed for the analysis of optimality conditions  in the
setting of  large neuronal populations and large output
signal-to-noise
(SNR) ratios \cite{r:brunelnadal,  r:mcdonnell08prl}.

In this paper  we examine the effect of vanishing signal amplitude on
the information transfer. We are motivated by the situation observed
in sensory neurons, which are in many cases known  to be responding
to weak stimuli intensities (relative to the external or internal
noise sources) \cite{r:bialekbits, r:deweeseopt, r:rieke}.
Information transfer  in channels subject to input cost constraints,
with implications to  low SNR  conditions, has also been of  interest
in information-theoretic literature \cite{r:verdu-capcost}.  In this
paper we employ a  different setting and examine information transfer
in channels with memory under vanishing stimulus amplitude constraint.
We explicitly consider the effect of channel memory, since many
realistic systems exhibit this property on various time scales, and
furthermore the presence of memory is known to enhance information
transfer in many cases \cite{r:cerf,  r:chacronprl, r:cover,
r:holden-cor}.  Finally, we apply the theory to calculate the effect
of memory on information transmission in a simple neuronal model
\cite{r:greenwoodprl, r:greenwoodpre}. This system exhibits the
stochastic resonance effect, which is commonly understood to be the
noise-induced enhancement of the system sensitivity to a \emph{weak}
signal \cite{r:mcdonnell-sr} (although  signal weakness
is not a necessary condition for the stochastic resonance
to occur \cite{r:stocksprl}).

Throughout this paper we assume \emph{discrete-time} setting, i.e., we
denote the consequent responses of a single stochastic neuronal unit
as a vector of random variables (discrete or continuous) $\mathbf
R=(\{R_i\}_{i=1}^{n})^\T$, where $i$ indexes the time and
$(\cdot)^{\T}$ denotes the transposition.  The response, $R_i=r_i$, is
invoked by stimulus, $\Theta_i=\theta_i$, where the stimulus course in
time is described by a $n$-dimensional vector of random variables
(r.v.) $\bm\Theta$.  We account for the \emph{memory} of the neuron,
so that  $R_i$ generally depends on  current, but also on past
stimulations and responses.  In the following we assume that the
neuronal model is realized by the stationary causal discrete-time
information channel with continuous input, fully described by the
conditional probability density function $f(\mathbf r|\bm\theta)$,
which factorizes as \cite{r:ash}
\begin{equation}
f(\mathbf r|\bm\theta)
=
\prod_{i=1}^n
f_i (r_i| 
\theta_i, \theta_{i-1}, \dots, \theta_1,
r_{i-1}, \dots, r_1).
\end{equation}
 In our setting we do not consider channel \emph{feedback}, i.e.,
dependence of current stimulus  on past responses. 

The two most well-known information measures, Fisher information (FI) 
and
Shannon's mutual information (MI), rely on $f(\mathbf r|\bm\theta)$.
The FI (matrix) is often employed as a measure of 
the efficiency of the population coding \cite{r:brunelnadal, 
r:sompolinsky-cor},
\begin{equation}
\mathbf J(\bm\theta|\mathbf R)=
\aver{[\nabla \ln f(\mathbf r|\bm\theta)]
[\nabla \ln f(\mathbf r|\bm\theta)]^\T}{\mathbf r|\bm\theta},
\label{eq:fi}
\end{equation}
where the gradient is with respect to $\bm\theta$, 
 and
$\aver{\cdot}{\mathbf r|\bm\theta}$ denotes averaging with respect to
$f(\mathbf r|\bm\theta)$. Throughout this paper we assume that
$f(\mathbf r|\bm\theta)$ is sufficiently continuous in $\bm\theta$, so
that the following regulatory conditions \cite{r:kay} hold
\begin{equation}
\int_{\mathbf R} \nabla
f(\mathbf r| \bm\theta)\,d\mathbf r=\mathbf 0,
\quad
\int_{\mathbf R} \nabla\nabla^{\T}
 f(\mathbf r| \bm\theta)\,d\mathbf r=\mathbf 0.
\end{equation}
FI imposes  limits on the precision of
$\bm\theta$ estimation from the responses, namely,
for the variance of any unbiased estimator of $\theta_i$ holds
$\var(\hat{\theta}_i)\geq [\mathbf J^{-1}(\bm\theta|\mathbf R)]_{ii}$
\cite{r:kay}.

MI is the fundamental quantity measuring information transfer in
channels \cite{r:cover}.
MI gives the degree of
statistical dependence between stimuli and responses and is
defined as
\begin{equation}
I(\bm\Theta;\mathbf R)= \aver{
\aver{\ln \frac{f(\mathbf
r|\bm \theta)}{p(\mathbf r)}}{\mathbf r|\bm\theta}
}{\bm\theta},
\label{eq:mi}
\end{equation}
where $p(\mathbf r)=\aver{f(\mathbf r|\bm \theta)}{\bm\theta}$
describes the marginal distribution of responses, and the averaging is
with respect to the distribution of stimuli, $\pi(\bm\theta)$, so that
MI is essentially property of the joint distribution of stimuli and
responses. The maximum value of MI per time step, taken over all
possible stimuli distributions, is
the information
capacity (or capacity rate), $\capa$, defined as
\cite{r:cover},
\begin{equation}
\capa= \lim_{n\rightarrow\infty} \max_{\pi(\bm\theta)}
\frac{1}{n} I(\bm\Theta;\mathbf R).
\label{eq:capa}
\end{equation}

FI is a local quantity in the sense that for some $\bm\theta_0$, 
$\mathbf J(\bm\theta_0|\mathbf R)$ takes into account stimuli from 
an infinitesimal neighbourhood of $\bm\theta_0$.
In other words, if we assume that FI is a real
quantity, i.e., something that can be measured and taken into account,
 then the stimuli from the neighbourhood of $\bm\theta_0$ 
have to be physically
present, which makes 
FI analogous to MI in the following sense.
Let the stimuli be restricted in amplitude, so that for some
$\bm\theta_0$ and $\Delta\bm\theta$ holds $
\bm\Theta\in [\bm\theta_0-\Delta\bm\theta,
\bm\theta_0+\Delta\bm\theta]$ and $\Delta\theta_i >0$.
We define a shifted r.v. 
 $\delta\bm\Theta$ as $\delta\bm\Theta= \bm\Theta- \bm\theta_0$ and
rewrite the MI from Eq.~(\ref{eq:mi}) in terms of r.v.
$\delta\bm\Theta\sim \pi(\delta\bm\theta)$ as
\begin{equation}
I(\bm\Theta;\mathbf R)=
\aver{\int_{\mathbf R}
\left[
\varphi(\mathbf r, \bm\theta_0+\delta\bm\theta)
- \psi(\mathbf r, \bm\theta_0+\delta\bm\theta)
\right]\,d\mathbf r
}{\delta\bm\theta},
\label{eq:mutinfcm}
\end{equation}
by further introducing 
\begin{eqnarray}
\varphi(\mathbf r, \bm\theta_0+\delta\bm\theta)&=&
f(\mathbf r|\bm\theta_0+\delta\bm\theta) \ln
f(\mathbf r|\bm\theta_0+\delta\bm\theta),
\\
\psi(\mathbf r, \bm\theta_0+\delta\bm\theta)&=&
f(\mathbf r|\bm\theta_0+\delta\bm\theta) \ln
\aver{f(\mathbf r|\bm\theta_0+ \delta\bm\theta)}{\delta\bm\theta}.
\end{eqnarray}
Now we consider the case of vanishing amplitude,
$\|\Delta\bm\theta\|\geq \|\delta\bm\theta\|\rightarrow 0$, 
and  expand
$I(\bm\Theta;\mathbf R)$ in Eq.~(\ref{eq:mutinfcm})  around
$\bm\theta_0$ in terms of $\delta\bm\theta$.
It can be shown 
\footnote{Extended manuscript is in preparation.} 
that,
\begin{equation}
\ln
\aver{f(\mathbf r|\bm\theta_0+ \delta\bm\theta)}{\delta\bm\theta}
\approx \ln f(\mathbf r| \bm\theta_0) 
+ \aver{\delta\bm\theta}{}^\T 
\frac{\nabla f(\mathbf r| \bm\theta_0)}{f(\mathbf r| \bm\theta_0)},
\end{equation}
where $\aver{\delta\bm\theta}{}= 
\aver{\delta\bm\theta}{\delta\bm\theta}$, and thus the Taylor
expansion of  $\psi\equiv\psi(\mathbf r,
\bm\theta_0+\delta\bm\theta)$,
is
\begin{eqnarray}
\psi &\approx&
f\ln f +\delta\bm\theta^{\T} \ln f \nabla f 
+\aver{\delta\bm\theta}{}^{\T} \nabla f+
\nonumber\\
&&+\frac{1}{2} \delta\bm\theta^{\T}\ln f \nabla\nabla^{\T}
f\,\delta\bm\theta+
\delta\bm\theta^{\T} \frac{\nabla f\nabla^{\T}
f}{f}\aver{\delta\bm\theta}{}+
\nonumber\\
&&+\frac{1}{2} \aver{\delta\bm\theta}{}^{\T} f \left[
\frac{\nabla\nabla^{\T} f}{f} -\frac{\nabla f\nabla^{\T} f}{f^2}
\right] \aver{\delta\bm\theta}{},
\end{eqnarray}
where  $f\equiv f(\mathbf r|\bm\theta_0)$ and $\nabla
f\equiv \nabla f(\mathbf r| \bm\theta)|_{\bm\theta=
\bm\theta_0}$. The analogous
expansion of $\varphi$ is straightforward.
By substituting the expansions into Eq.~(\ref{eq:mutinfcm})
the zeroth- and first-order terms cancel
and what remains can be written in terms of FI matrix evaluated at
$\bm\theta=\bm\theta_0$, by employing
 $\mathbf J(\bm\theta_0|\mathbf R)=[\mathbf
J(\bm\theta_0|\mathbf R)]^\T$,
as
\begin{equation}
I(\bm\Theta;\mathbf R)
\approx
\frac{1}{2}
\aver{ 
\left[\delta\bm\theta-\aver{\delta\bm\theta}{}\right]^\T
\mathbf J(\bm\theta_0|\mathbf R)
\left[\delta\bm\theta-\aver{\delta\bm\theta}{}\right]
}{\delta\bm\theta},
\label{eq:mi-s1}
\end{equation}
and after taking the expectation
\begin{equation}
I(\bm\Theta; \mathbf R) 
\approx
\frac{1}{2} \tr\left[
\mathbf J(\bm\theta_0|\mathbf R)
\mathbf C_{\bm\Theta}
\right],
\label{eq:weakcapa}
\end{equation}
where $\mathbf C_{\bm\Theta}$ is the
covariance matrix of $\bm\Theta$ and $\tr(\cdot)$ is the matrix trace.
Eq.~(\ref{eq:weakcapa}) holds for a broad class of channels with
memory, both biologically-inspired and artificial, and represents
the main result of this paper. 

Next we concentrate on the interpretation and some immediate
implications of Eq.~(\ref{eq:weakcapa}).
First, the information capacity from Eq.~(\ref{eq:capa}) follows
readily from Eq.~(\ref{eq:weakcapa}): FI matrix is the property of the
neuronal model, so the stimulus properties are represented by $\mathbf
C_{\bm\Theta}$. Maximizing $I(\bm\Theta; \mathbf R)$ thus corresponds
to extremizing the values of $[\mathbf C_{\bm\Theta}]_{ik}$ for which
the corresponding elements $[\mathbf J(\bm\theta_0|\mathbf R)]_{ik}$
are non-zero (with appropriate sign).
 E.g., for a memoryless channel, $f(\mathbf r|\bm\theta)
=\prod_{i=1}^n f_i (r_i| \theta_i)$, so the FI matrix is diagonal with
elements $[\mathbf J(\bm\theta_0|\mathbf R)]_{ii}= J(\theta_0|R)$
(omitting the index $i$ due to channel stationarity). The capacity is
thus achieved by maximizing the variance of the amplitude-constrained
stimulus, so the capacity-bearing 
distribution is  realized by two
equiprobable probability masses located at the interval extremes, and
\begin{equation}
\capa = \frac{1}{2} (\Delta\theta)^2 J(\theta_0|R),
\end{equation}
a result obtained by different means in \cite{r:verdu-capcost}.
Generally, $I(\bm\Theta; \mathbf R)\rightarrow 0$ as the stimulus
amplitude vanishes. It is thus advantageous to
introduce the MI (and capacity) per maximum stimulus power, i.e., 
$\bar{I}(\bm\Theta;\mathbf R)= I(\bm\Theta;\mathbf
R)/\|\Delta\bm\theta\|$,
so for the memoryless channel $\bar{\capa}= J(\theta_0|R)/2$, as 
obtained in \cite{r:verdu-capcost}.
While the previously mentioned asymptotics of MI in terms of FI
\cite{r:brunelnadal,  r:mcdonnell08prl} deals with the
low-noise limit  of information transmission
(i.e., large neuronal populations),
 Eq.~(\ref{eq:weakcapa}) describes the opposite
``large-noise'' limit situation.

In the following we apply Eq.~(\ref{eq:weakcapa})
 on the classical
McCulloch-Pitts (MP) neuronal model,
 accounting for the memory of the noise
component. Memoryless variant of the MP model has been sucesfully
employed in describing the stochastic resonance effect in
electrosensory neurons of paddlefish \cite{r:greenwoodprl}, and
further analyzed in detail in \cite{r:greenwoodpre, 
r:langreen-bc}.
The MP model is based on thresholding of the stimulus (corrupted
by an additive noise $\mathbf X$), so that the discrete-valued
response in time-step $i$ is
\begin{equation}
R_i= U(\theta_i+ X_i-a),
\label{eq:mcpn}
\end{equation}
where $a$ is the threshold, $U(\cdot)$ is the Heaviside step function
and $\theta_i\in [-\Delta\theta+\theta_0, \theta_0+\Delta\theta]$ for
all $i$. The
occurrence of action potential at time $i$ is  indicated by $R_i=1$.
In the following we consider the noise r.v. 
$\mathbf X=\{X_1, \dots, X_n\}^\T$
to be identically distributed but \emph{dependent}, which provides the
memory effect for the MP neuron. For simplicity, we assume in the
following that $\mathbf X\sim p(\mathbf x)$ is gaussian with
covariance matrix $[\mathbf C_{\mathbf X}]_{ik}= \sigma^2
\varrho_{ik}$, where $\varrho_{ik}= \corr(X_i, X_k)$ is the serial
correlation coefficient.
Obviously, since $U$ is not invertible, any simple form of dependence
in the noise (such as first order Markov) is not preserved in the time
sequence of responses.  Generally, the full joint distribution of
$\mathbf R$ is required, which means evaluation of  $n$-dimensional
gaussian integrals, which may not be numerically stable. 
In other words, the joint conditional probabilities 
$\pr{\mathbf R|\bm\theta}$ are generally not tractable for
reasonable values of $n$.
The idea is to substitute the full and untractable log-likelihood,
$\ell(\bm\theta|\mathbf r)= \ln f(\bm r|\bm \theta)$, 
 with a  computable pseudo-log-likelihood \cite{r:molenberghs},
$\ell^{(P)}(\bm\theta| \mathbf r)$,  neglecting some high-order
dependencies, i.e.,
\begin{equation}
\ell^{(P)}(\bm\theta|\mathbf r)
=
\sum_{q} \ell^{(P)}_q(\bm\theta| \mathbf r),
\end{equation}
where $\ell^{(P)}_q(\bm\theta| \mathbf r)$ are ``computable''
partitions.
Here we concentrate on a variant of the
 second-order pseudo-log-likelihood,
$\ell^{(P)}(\bm\theta|\mathbf r)=\ell_2(\bm\theta| \mathbf r)$,
based on \emph{pairwise} dependence \cite{r:coxreid}
\begin{equation}
\ell_2(\bm\theta| \mathbf r)=
\sum_{i=2}^{n}
\sum_{k=1}^{i-1} \ln \pr{R_i, R_k| \bm\theta}- (n-2)
\sum_{i=1}^n  \ln \pr{R_i|\bm\theta},
\label{eq:lpair} 
\end{equation}
The advantage of $\ell_2$ is, that most of the involved integrals can
be expressed in a semi-closed form for the gaussian noise.
The problematics of replacing 
$\ell(\bm\theta|\mathbf r)$ by
$\ell_2(\bm\theta|\mathbf r)$ for non-Markov models has been
investigated recently in statistical literature
\cite{r:coxreid, r:nottryden}.
The marginal probability $P_1$ of 
$R_i=1$ (crossing the threshold) is independent of $i$ due to
stationarity,  and since $R_i\in\{0,1\}$, we can write
$\pr{R_i|\theta_i}=
r_i P_{1}+ 
(1-r_i) (1-P_1)
$, where
\begin{eqnarray}
P_1= \frac{1}{2} \left[
1-\erf\left(\frac{a-\theta_i}{\sqrt{2} \sigma}\right)
\right]
\label{eq:mcpmarg}
\end{eqnarray}
by evaluation of the gaussian integral and $\erf(\cdot)$ is the error
function. Similarly, for the bivariate
joint response probability holds
\begin{eqnarray}
\pr{R_i, R_k| \theta_i, \theta_k} =
r_i r_k P_{11} + r_i (1-r_k) P_{10} \nonumber\\
+ (1-r_i) r_k P_{01} + (1-r_i) (1-r_k) P_{00},
\end{eqnarray}
where $P_{mn}= P_{mn}(\theta_i, \theta_k)$ 
is the probability of $R_i=m, R_k=n$, so $\sum_{m,n}
P_{mn}=1$.
Note, that $P_{11}+P_{01}$ is the marginal
probability of $R_k=1$, and  $P_{11}+P_{10}=P_{1}$ is the
marginal probability of $R_i=1$.
Eq.~(\ref{eq:mcpmarg}). These symmetries and Eq.~(\ref{eq:mcpmarg})
give
\begin{eqnarray}
P_{11} 
&=& \int\limits_{0}^{\infty}\frac{1}{2} \left[
1+\erf\left(
\frac{\theta_i-a+
(a-\theta_k+y)\varrho_{ik}}{\sigma\sqrt{2-2\varrho^2_{ik}}}
\right)
\right] \times\nonumber\\
&\times&\phi(y-\theta_k+a)\,dy,
\label{eq:P11}
\\
P_{01}
&=&\frac{1}{2} \left[
1-\erf\left(\frac{a-\theta_k}{\sigma\sqrt{2}}\right)
\right] -P_{11},
\\
P_{10}&=&\frac{1}{2} \left[
1-\erf\left(\frac{a-\theta_i}{\sigma\sqrt{2}}\right)
\right]- P_{11}, 
\\
P_{00}&=& 1 - P_{11}-P_{01} -P_{10}.
\end{eqnarray}
where $\phi(\cdot)$ is the probability density function of a gaussian
r.v.
with zero mean and variance equal to $\sigma^2$ (note that $P_{mn}$ are
functions of $\theta_i, \theta_k,a,\sigma$ and $\varrho_{ik}$).
\begin{figure}[t]
\includegraphics{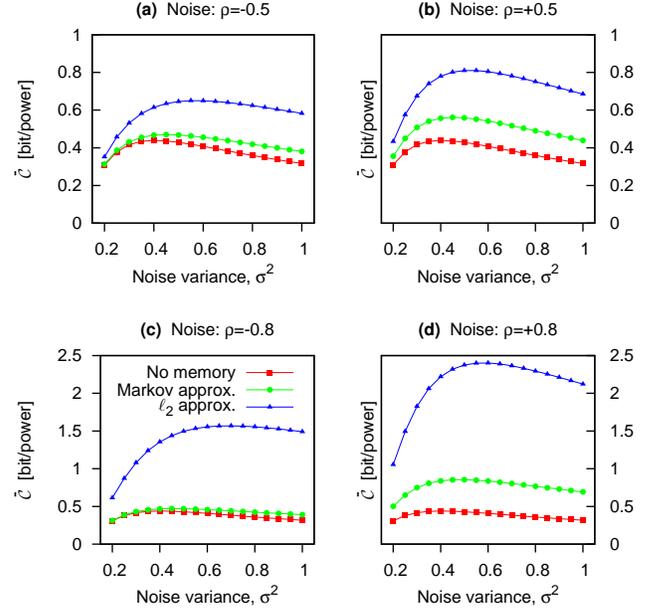}
\caption{Information capacity  (in bits) per vanishing
stimulus power of McCulloch-Pitts neuronal model with memory.  The
noise is a gaussian AR(1) process with first-order serial correlation
$\varrho$ and variance $\sigma^2$. Stimulation parameters are:
$\theta_0=0$ and threshold $a=1$.  Three situations are shown:
\emph{no memory} (se also \cite{r:greenwoodpre, r:langreen-bc},
corresponds to $\varrho=0$), \emph{Markov} (assuming the first-order
Markov structure of responses) and \emph{pseudo-log-likelihood},
$\ell_2$-approximation to the true situation, estimated for $n=100$.
The memory of the neuron enhances its information capacity, by
reducing the disruptive power of the noise. Note, that positive noise
correlations increase capacity more than negative ones.
\label{fig:fig1}}
\end{figure}
The FI
matrix will have generally all elements non-zero, and its
approximation by $\ell_2$ is
\begin{eqnarray}
[\mathbf J(\bm\theta|\mathbf R)]_{ik}
&=&-\sum_{r_1, \dots, r_n} \pd{^2 \ell_2(\bm\theta|\mathbf R)}
{\theta_i\,\partial\theta_k}\times\nonumber\\
&\times&\pr{[R_1=r_1, \dots, R_n=r_n]|\bm\theta},
\end{eqnarray}
where the sum is over all possible $n$-dimensional 
vectors, consisting  of $0$'s and~$1$'s.
Due to
particular form of $\ell_2(\bm\theta|\mathbf T)$, however, things are
a lot simpler, although  details of the following calculations will be
published elsewhere. For the off-diagonal, $i\neq k$,
and diagonal elements evaluated at
$\theta_i=\theta_k=\theta_0$ holds
\begin{eqnarray}
\mathbf J(\bm\theta|\mathbf R)_{ik}&=&
 \gamma\left(
\theta_0, a, \sigma, \varrho_{ik}, \tilde P_{11}, \tilde\phi_{01}
\right),\\
\mathbf J(\bm\theta|\mathbf R)_{ii}&=&
\omega\left(\theta_0, a, \sigma, \varrho_{ii},
\tilde P_{11}\right),
\end{eqnarray}
where  $\gamma(\cdot)$ and $\omega(\cdot)$ are
 complicated (but tabulated) functions of the
indicated parameters, and 
\begin{eqnarray}
\tilde P_{11} &=& P_{11}(\theta_0, \theta_0),
\label{eq:ptilde}
\\
\tilde\phi_{01}&=&\left.
\pd{}{\theta_k} P_{11}(\theta_0, \theta_k)
\right|_{\theta_k=\theta_0}.
\label{eq:phi01}
\end{eqnarray}
Employing Eq.~(\ref{eq:weakcapa}) gives the covariance matrix of the
optimal stimulation as
\begin{equation}
[\mathbf C_{\bm\Theta}]_{ik}= (\Delta\theta)^2 
\,\sgn\left([\mathbf
J(\bm\theta_0|\mathbf T)]_{ik}\right),
\label{eq:optincov}
\end{equation}
where $\sgn(\cdot)$ is the signum function. The capacity rate per
vanishing stimulus power is then
\begin{equation}
\bar\capa=
\lim_{n\rightarrow\infty} \frac{1}{2n}
\sum_{i,k}
|[\mathbf J(\bm\theta_0|\mathbf T)]_{ik}|.
\label{eq:capamcpn}
\end{equation}
Fig.~\ref{fig:fig1} shows how the memory of the neuron enhances
its information capacity (shown as a capacity per vanishing stimulus
power). We assumed that the noise r.v. $\mathbf X$
 is modelled by
the AR(1) gaussian discrete-time process with first-order correlation
$\varrho$, so that $[\mathbf C_{\mathbf X}]_{ik}=\sigma^2
\varrho^{|i-k|}$. The enhancement is compared to the already
investigated $\varrho=0$ case (no memory) \cite{r:greenwoodpre,
r:langreen-bc}), which exhibits the effect of stochastic resonance as
the variance of the noise increases.
The information transferred increases with memory, since the noise
correlations effectively reduce its ``corrupting'' power (once the
stimulus statistics is properly matched to the noise structure, as
shown by Eq.~(\ref{eq:weakcapa})).
The \emph{no memory}
values are identical in all cases, since the noise correlations are
ignored.
 Besides the 
$\ell_2$-approximation, the first-order Markov approximation is also
shown, obtained by setting $n=2$ in Eq.~(\ref{eq:lpair}). For
Markov approximation the information capacity is lower, since
the neuron employs only current and immediately
preceeding response value in the decoding, neglecting the
possibilities of the essentially infinite-memory of the MP neuron.
Additional numerical calculations show,
that even small noise correlations ($\varrho\approx 0.2$)
increase the capacity rates of the MP neuron by approx. $15\,\%$
(not shown in Fig.~\ref{fig:fig1}).


Our results lead us to comment on the optimality of information
transfer in real neurons. While the efficient coding hypothesis relies
on the maximum information transfer, one should keep in mind, that
from the information-theoretic perspective the coding-decoding
operations are an integral part of the information transmission
process.  First, it is well known \cite{r:cover}, that for some
channels the optimal decoding process can be a very complex task --
i.e., employing \emph{all} the responses obtained so far, as
illustrated in this paper on a relatively simple example of the MP
neuron with memory. Since the nervous system is assumed to respond to
spike trains in real time \cite{r:holden-cor}, it is questionable
that real neurons try to achieve the true capacity and additional
costs must be taken into account \cite{r:laughlin-met}.  Second, the
discrete, or impulse-like, character of capacity-bearing stimulation
is not limited only to vanishing stimulus amplitudes.  This phenomenon
occurs in most channels examined in literature so far (with
power-constrained AWGN channel, and low-noise limit channels being the
only known exceptions) \cite{r:chancap}.  Another possible problem
connected with the usage of a continuously varying stimulus is, that
the complete specification of particular $\theta$ requires infinite
amount of information, while real neurons probably do not strive for
precise specification of $\theta$. 

\begin{acknowledgments}
This work was supported by 
 AV0Z50110509 and  Centre for
Neuroscience LC554.
\end{acknowledgments}


%

\end{document}